\documentclass[journal,comsoc]{IEEEtran}
\usepackage{cite}
\usepackage{amsmath,amssymb,amsfonts}
\usepackage{graphicx}
\usepackage{textcomp}
\usepackage{xcolor}
\usepackage{subfigure}
\usepackage{threeparttable}
\usepackage{mathrsfs}
\usepackage{url}
\usepackage{mdwtab}
\usepackage{booktabs}
\usepackage{array}
\usepackage{mdwmath}
\usepackage{eqparbox}
\usepackage{multirow}
\usepackage{color}
\usepackage{threeparttable}
\usepackage{array}
\usepackage{booktabs}
\usepackage{textcase}
\usepackage{siunitx}
\usepackage{hyperref}
\usepackage{makecell}

\usepackage{pifont}
\newcommand{\cmark}{\ding{51}}

\begin{document}


\title{\textsc{DefenGraph}: Knowledge Graph-Enhanced LLMs for Blue Team Cyber Defense}


\author{
Zhen Wang,
Kristen Moore,
Qin Wang,
Guangsheng Yu,
Minjune Kim,
Diksha Goel,
Gang Li,
Ahmed Ibrahim,
Ahmad Mohsin,
and Helge Janicke%
\thanks{Zhen Wang, Kristen Moore, Qin Wang, Minjune Kim, and Diksha Goel are with CSIRO's Data61, Australia.
(e-mail: \{jeff.wang, kristen.moore, qin.wang, minjune.kim, diksha.goel\}@data61.csiro.au)}%
\thanks{Guangsheng Yu is with University of Technology Sydney, Australia.
(e-mail: guangsheng.yu@uts.edu.au)}%
\thanks{Ahmad Mohsin, Ahmed Ibrahim and Helge Janicke are with Edith Cowan University, Australia.
(e-mail: \{a.mohsin, a.ibrahim, h.janicke\}@ecu.edu.au)}%
\thanks{Gang Li is with Deakin University, Australia.
(e-mail: gang.li@deakin.edu.au)}%
}

\markboth{Manuscript submitted to IEEE Transactions on Information Forensics and Security}%
{Shell \MakeLowercase{\textit{et al.}}: A Sample Article Using IEEEtran.cls for IEEE Journals}


\maketitle

\begin{abstract}

Large Language Models (LLMs) show promise for supporting decision-making in cybersecurity, but their reliability in high-stakes, time-evolving environments remains limited due to hallucinations, poor temporal reasoning, and shallow grounding in system context. 
We introduce \textsc{DefenGraph}, an LLM-driven assistant designed to support human defenders during cybersecurity incidents. \textsc{DefenGraph} improves contextual reasoning by integrating a dual-layer Static-Dynamic Knowledge Graph (KG) with graph-based path retrieval, LLM-driven contextual filtering, and reasoning-based re-ranking. The framework grounds LLM outputs in both long-term domain knowledge and evolving event context, enabling faithful and temporally aware decision support. We evaluate \textsc{DefenGraph} in a cyber defense setting using knowledge graphs constructed from heterogeneous security artifacts, including SIEM alerts, system topology, attacker behaviors, and prior defensive actions. The evaluation uses data collected during live Red vs. Blue team cyber range exercises simulating attacks on critical infrastructure, which generate realistic and noisy datasets reflecting real-world defender workflows and system dynamics.
Evaluations across four prevalent LLMs show that \textsc{DefenGraph} sets a new state-of-the-art: on GPT-4o it boosts reasoning-recall from 61.45\% to 73.49\% and ticket-action recall from 52.17\% to 72.46\% (precision 24.49\%~$\rightarrow$~29.24\%), with similar gains on LLaMA-3 (46.99\%~$\rightarrow$~61.45\%), DeepSeek-R1 (45.78\%~$\rightarrow$~56.63\%) and QWen-3 (51.81\%~$\rightarrow$~59.04\%), while surfacing up to 50 correct defense actions versus 36 for the next-best baseline and holding fault rates steady.

\end{abstract}


\section{Introduction}

Large Language Models (LLMs) have demonstrated strong capabilities in question answering, and multi-step reasoning~\cite{sui2025stopoverthinkingsurveyefficient,ma2025soksemanticprivacylarge}. However, their reliability remains a critical concern in high-stakes environments, such as cybersecurity, where trustworthiness and domain-grounded reasoning are essential~\cite{Zhang2025, 10.1145/3769676, 10944629}. Cyber defenders must make fast, high-consequence decisions based on incomplete and evolving information. LLMs often fall short in these settings due to hallucinations, difficulty interpreting time-sensitive alerts, and a lack of grounding in domain-specific system context~\cite{10.1145/3703155,arikkat2024intellbotretrievalaugmentedllm}.

To explore how LLMs might better support defenders, we conducted a series of Red vs. Blue team (attacker vs. defender) training events in a cyber range simulating real-world critical infrastructure. These exercises generated rich, multi-modal data, including Security Information and Event Management (SIEM) alerts (via Wazuh\footnote{Wazuh is a free and open source Security Information and Event Management (SIEM) platform used for threat detection and prevention.}), structured incident response tickets (via Cydarm\footnote{Cydarm is a cybersecurity tool used in this research for incidents management within Cyber Range. \url{https://www.cydarm.com/}}), attacker activity traces, and system state~\cite{10482206,s25030870}.

While Retrieval-Augmented Generation (RAG) has improved LLM reliability by grounding outputs in external text, it remains brittle in structured domains~\cite{karpukhin-etal-2020-dense}. Traditional RAG relies on surface-level or embedding-based similarity, often retrieving irrelevant documents that confuse LLM reasoning~\cite{fan2024survey}. For instance, we observed cases from our dataset where a RAG system retrieved an incident response ticket for a Windows-based authentication error in response to a Linux-based SIEM alert, leading to flawed mitigation advice (see Fig.~\ref{fig:rag_passage_example}). 

\begin{figure}[t]
\scriptsize
\centering
\fbox{%
  \begin{minipage}{0.45\textwidth}
    \textbf{Input Wazuh Alert:}
    \begin{itemize}[leftmargin=*, itemsep=0pt]
        \item \textbf{agent\_name:} RSLGB\_Parter
        \item \textbf{title:} System user successfully logged to the system.
        \item \textbf{full\_log:} rlsgb su[15784]: \textbf{pam\_unix(su:session)}: session opened for user \textbf{nobody} by (uid=0)
    \end{itemize}
    \rule{\linewidth}{0.2pt}
    \textbf{Retrieved Ticket: ZF4C66}
    \begin{itemize}[leftmargin=*, itemsep=0pt]
        \item \textbf{Event Type:} Successful Logon
        \item \textbf{Source:} \textbf{Microsoft-Windows-Security-Auditing}
        \item \textbf{Logon Type:} 3 (Network Logon)
        \item \textbf{Account Name:} \textbf{ANONYMOUS LOGON}
        \item \textbf{Domain:} NT AUTHORITY
        \item \textbf{Workstation:} CONTRACTADMIN
        \item \textbf{Source IP:} 192.168.126.118
        \item \textbf{Auth Protocol:} NTLM V1
    \end{itemize}
    \textit{This Windows ticket describes an anonymous logon from CONTRACTADMIN using NTLM V1, mismatched with the Linux alert above.}
  \end{minipage}
}
\caption{Top: Wazuh alert for a suspicious login to a Linux server. Bottom: RAG retrieves an unrelated Windows ticket, missing key Linux context (e.g., \textbf{nobody} user, \textbf{pam\_unix} session).}
\label{fig:rag_passage_example}
\end{figure}

\begin{figure}
    \centering
    \includegraphics[width=0.9\linewidth]{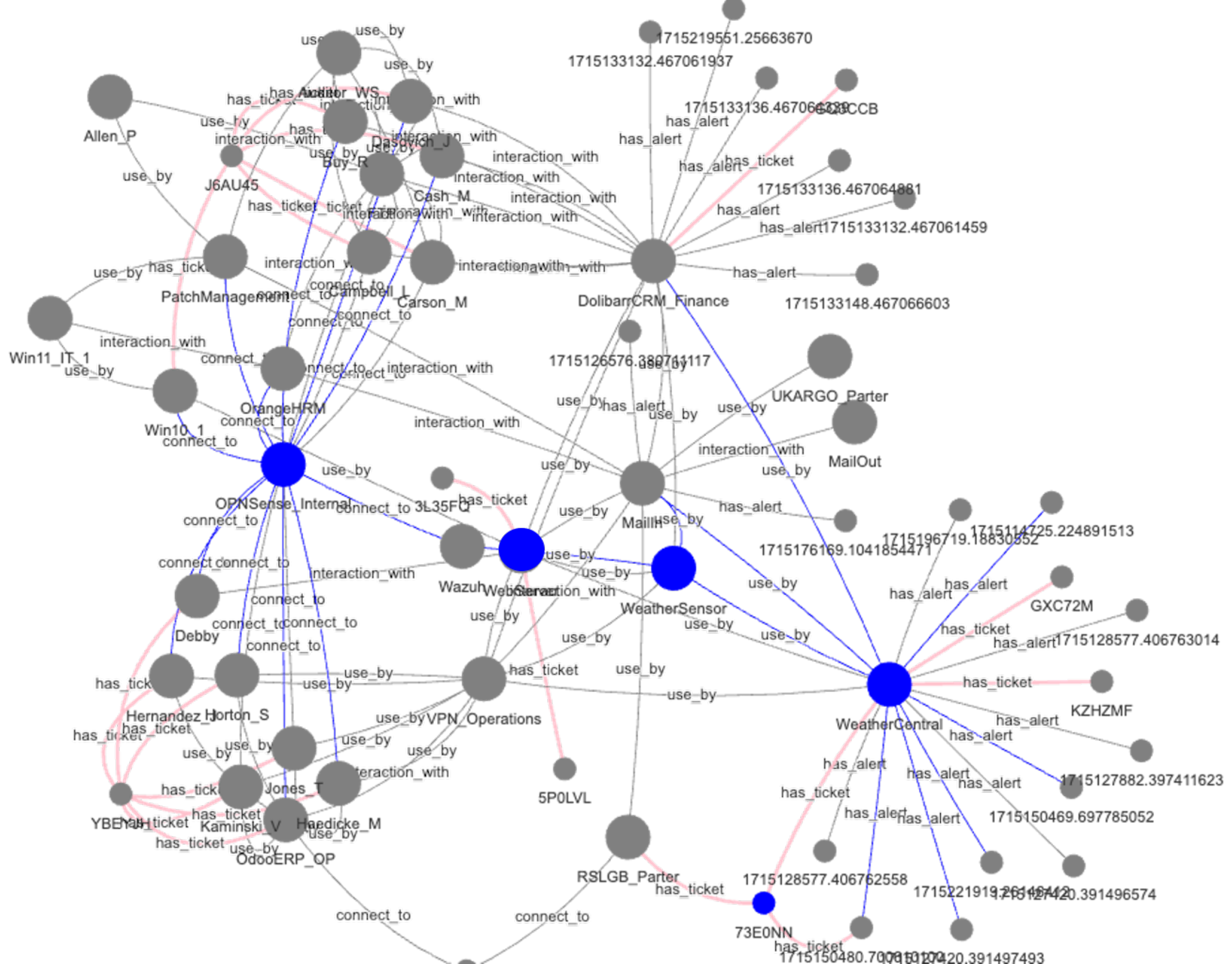}
    \caption{Complex and noisy context in Cybersecurity defense. \textbf{Blue nodes} represent critical information for generating cyber-defense activities. \textbf{Grey nodes} denote non-essential or noisy information.}
    \label{fig:noisy_context}
\end{figure}

Knowledge Graphs (KGs) offer a structured way to represent entities and their relationships, enabling contextual reasoning over system events~\cite{li2025context,an2024make}.  In cybersecurity, KGs have emerged as a promising solution for integrating heterogeneous data sources with interpretable schema. For example, Garrido et al~\cite{garrido2021machine} demonstrated that KGs can enhance intrusion detection by combining structured information with link prediction techniques. Their results showed that alerts derived from KGs reasoning were better calibrated and more interpretable than traditional methods. Traditional Knowledge Graph Retrieval-Augmented Generation (KG-RAG)~\cite{edge2024localglobalgraphrag, zhang-etal-2022-subgraph} typically enhances LLM performance by leveraging neighbourhood structure and precise node retrieval to provide richer contextual signals. However, this assumption does not hold in our cyber-defense setting. Cyber-defense data has a very low signal-to-noise ratio and contains large volumes of meaningless or misleading logs, such as user authentication failures caused simply by mistyped passwords. As shown in Fig.~\ref{fig:noisy_context}, while the blue nodes represent relevant information, the surrounding context is highly complex and contains extensive noisy and loosely related grey-color nodes which KG-RAG methods fail to filter out.

In this work, we explore an LLM-based contextual reasoning filter within KG-RAG, where relevant paths from both static and dynamic knowledge graphs are extracted by our semantic path selection and enhanced by an LLM-based contextual filter and a reasoning re-ranker. This enables more precise knowledge to be incorporated into the LLM’s input and better contextualization of alerts and alignment with system state and past defensive actions.

We present \textsc{DefenGraph}, a dual-layer KG-RAG framework that combines structured knowledge representations with LLM-based generation to support cyber defence. The graphs capture both historical system knowledge and real-time interactions, enabling the LLM to reason over temporally evolving system states and historical defense knowledge. Built from our cyber range exercises, these temporal knowledge graphs are enriched with artifacts such as Wazuh alerts and incident response tickets, etc. \textsc{DefenGraph} moves beyond static rule sets and keyword-based retrieval by retrieving relational paths that reflect relevant context for a new alert. These paths are processed through an LLM-based contextual filter and a reasoning re-ranker, enhancing retrieval quality by emphasizing semantically and temporally aligned information. The resulting paths are then embedded into the LLM prompt to guide generation of mitigation strategies that are situationally appropriate and aligned with the evolving threat landscape. 

To demonstrate the effectiveness of our approach in real-world cyber-defense scenarios, we evaluate \textsc{DefenGraph} using four leading LLMs (GPT-4o~\cite{DBLP:journals/corr/abs-2303-08774}, LLaMA 3.1~\cite{grattafiori2024llama3}, DeepSeek R1~\cite{deepseekai2025deepseekr1incentivizingreasoningcapability} and QWen 3~\cite{qwen3}) on a corpus of real alerts and defense tickets, comparing it to state-of-the-art RAG and KG-RAG baselines. Our results show that \textsc{DefenGraph} improves over both Hybrid RAG~\cite{yuan2024hybridragcomprehensiveenhancement} and GraphRAG~\cite{edge2024localglobalgraphrag} augmented LLMs in generating faithful, context-aware responses aligned with human-authored mitigation strategies. 

Our main contributions are as follows: 

\begin{itemize}
    \item We propose \textsc{DefenGraph} (\S\ref{sec:defengraph}), a temporal KG-RAG framework for generating LLM-based contextual reasoning strategies grounded in real-world cyber context.
    
    \item We construct and release the \textbf{first} comprehensive cybersecurity KG dataset (\S\ref{sec:acdc_dataset}), built from \textbf{real-world Red/Blue team exercises} conducted in our cyber range that replicates critical infrastructure, combining IT/IoT/Operational technology (OT) components in a realistic maritime port scenario.

    \item We demonstrate that \textsc{DefenGraph} enhances the faithfulness and interpretability of LLM outputs across multiple models, yielding defense recommendations that match Blue Team ground truth with markedly higher fidelity, lifting reasoning recall by 19.6\%, ticket-action recall by 20.3\%, and precision by 19.4\%.
\end{itemize}

The organization of the remaining paper is as follows: \S\ref{sec:acdc_dataset} introduces the ACDC cyber range dataset and the key data/labeling challenges it poses for defender decision support; \S\ref{sec:defengraph} presents the proposed DefenGraph framework and its static–dynamic KG construction and retrieval-to-generation pipeline; \S\ref{sec:exp} reports the experimental setup, baselines, evaluation metrics, and results across multiple LLMs; \S\ref{sec:rw} reviews related KG-based cybersecurity research; and \S\ref{sec:conclusion} concludes the paper.

\section{ACDC Cyber Range Dataset and Challenges}
\label{sec:acdc_dataset}

To address the gap between offensive and defensive cybersecurity research, we collaborated with an industry partner\footnote{Canberran cyber security experts, Penten. \url{https://canberra.com.au/business/investing-in-canberra/success-stories/success-story-penten} through Australian CSCRC project \url{https://cybersecuritycrc.org.au/other-projects/}.} to build a real-world dataset evaluating how LLMs can support blue teams. The dataset represents one of the most comprehensive, operationally realistic collections of cyber-defense data produced in a controlled research environment ACDC Cyber Range. 

The ACDC Cyber Range includes over 55 IT and OT systems and full architecture details are in Supplementary V. It collected attack and defense dynamics in a simulated maritime port environment and focuses on three interconnected scenarios representing real-world cybersecurity challenges in IT and operational technology (OT) systems:

\begin{itemize}
\item \textbf{IoT depth sensors attack.} Targets the integrity of tide monitoring systems.
\item \textbf{Conveyor belt attack.} Explores threats to OT components managing bulk material loading, focusing on programmable logic controllers (PLCs) and human-machine interfaces (HMIs).
\item \textbf{Train supply chain attack.} Highlights threats to autonomous train logistics via PLC-SCADA links.
\end{itemize}

It integrates heterogeneous data sources generated during a live, 5-day cybersecurity exercise involving 6 professional Red Team members and 8 Blue team members. The dataset captures interactions across the full cyber-defense lifecycle activity, SEMI signals (Wazuh alerts), team communication (Slack\footnote{We integrated Wazuh alerts into Slack channel used by Blue Team for communication, alert automation, and prioritization with \textsc{DefenGraph}.}), and incident-response workflows (Cydarm) which offer an unprecedented opportunity to study cyber operations.

Collecting this dataset required extensive technical coordination across sensors, SIEM systems, communication platforms, and human operators. Unlike simulated datasets or synthetic log collections, the ACDC dataset preserves the true friction, noise, and ambiguity of real-world cyber defense, including incomplete information, misaligned tooling, and human decision processes. The result is a dataset that is both extremely challenging to curate and highly valuable to the research community.

The final dataset consists of over 8.2M firewall log lines, 6.1M+ Wazuh alerts, 326 incident response tickets, 4.7k+ Red Team terminal activity, and 2000+ lines of Slack communication, all synchronized to a shared timeline that captures the evolving attack–defense dynamics. To generate the ground-truth attack-alert–ticket triples, expert validation was required. Because blue team members manually created incident-response tickets, establishing accurate triples mappings was challenging. After experts' review, we obtained 55 confirmed attacks–alerts-ticket triples, along with additional benign alerts to simulate real-world noise. 

When processing this dataset, we encountered the following data-processing challenges:

\smallskip
\noindent\textbf{Unstructured Free Text Data.}
A significant portion of the dataset, such as logs, chat messages, and server records, exists in unstructured formats. This lack of structure complicates data extraction, analysis, and integration, particularly for identifying relationships and correlations between dataset components.

\smallskip
\noindent\textbf{Misalignment Between Wazuh Alerts and Incident Response Tickets.}
Wazuh servers monitor high-volume log sources and rely on broad, generic detection rules. As a result, automatically generated Wazuh alerts often contain false positives or redundant information. Meanwhile, a full attack could involve multiple stages from the red teams such as port and service enumeration, password brute-forcing, email phishing, and more, which involves a series of Wazuh alerts. Therefore, Blue Team members must identify those suppicious alerts and link them together to understand the broader context. As a result, Blue Team members manually create incident response tickets to document defense actions with contextual judgment and each ticket may contain multiple related alerts. Note that some industry tools may help the alert-ticket linkage automatically, but this is beyond the scope of this paper. So the key issues include: 

\begin{itemize}
    \item \textbf{One-to-many mapping.} A single incident response ticket may link to multiple Wazuh alerts due to manual ticketing and alert redundancy. 

    \item \textbf{Duplicate alerts.} Repetitive or widespread issues can trigger multiple alerts, all addressed by one ticket, complicating alert-ticket matching.

    \item \textbf{No alert match.} Some tickets are created proactively without any corresponding alert, e.g., default password changes made before an anomaly is detected.

    \item \textbf{Irrelevant alerts.} Not all alerts are meaningful; false positives or benign events may have no connection to ticket actions.
\end{itemize}

\smallskip
\noindent\textbf{Red Team Activities and Wazuh Alerts Misalignment.}
During the exercises, the Red Teams activities in the Terminal shell (command line) were recorded. The Red Team's activities do not have a one-to-one correspondence with Wazuh alerts. This lack of precise mapping complicates efforts to validate and analyze the relationship between offensive actions and defensive responses.

\section{\textsc{DefenGraph} Framework}
\label{sec:defengraph}

\begin{figure*}[t] 
    \centering
    \includegraphics[width=\linewidth]{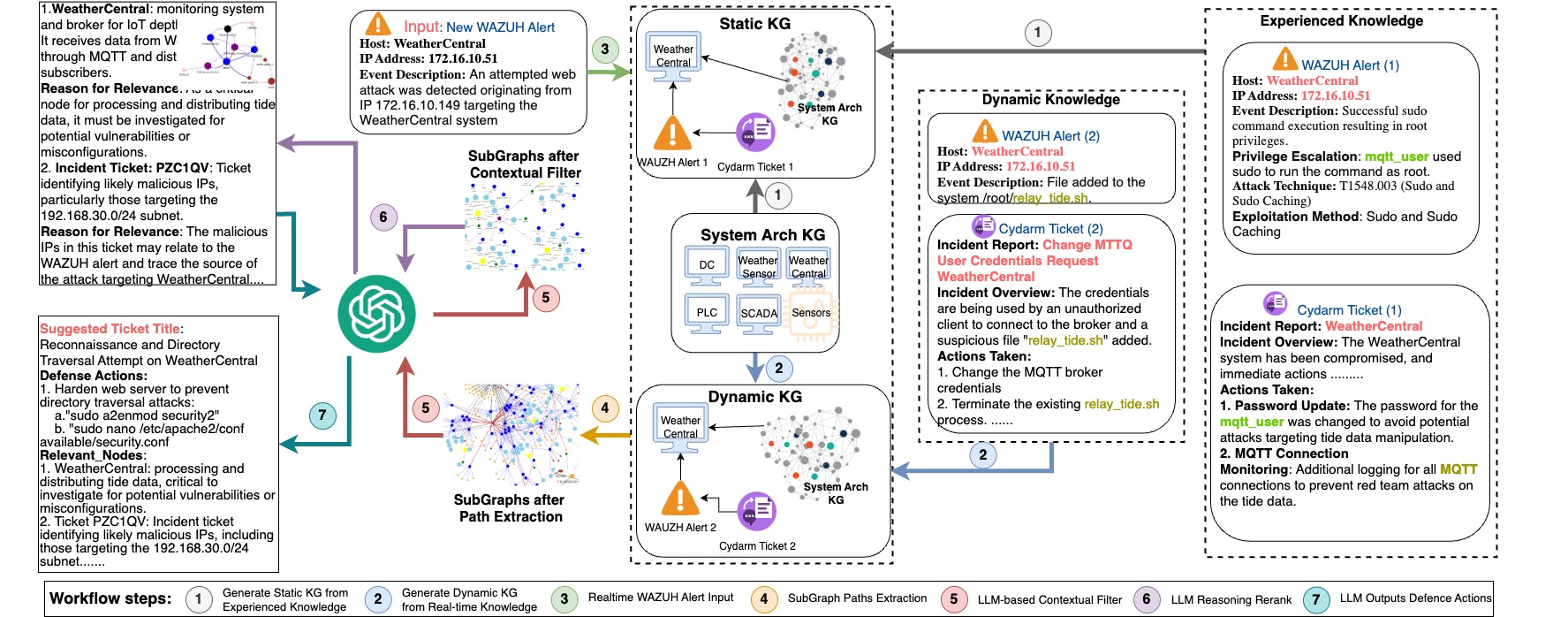} 
\caption{\textbf{System overview of \textsc{DefenGraph}}. Historical alerts and tickets are structured into a static KG, while new Wazuh alerts provide dynamic knowledge. The system KG encodes the host and service topology. A static–dynamic KG is built by fusing historical and real-time data. Given a new alert, the system extracts subgraphs, ranks reasoning paths with an LLM, and generates context-aware defense actions aligned with Blue Team practices.}
    \label{fig:defengraph}
\end{figure*}

To support reliable LLM-driven decision making in dynamic cyber environments, we proposed \textsc{DefenGraph}, a framework that augments LLMs with temporal knowledge graphs to generate context-aware mitigation strategies in response to SIEM alerts. \textsc{DefenGraph} is designed to help defenders reason over both static system structure and evolving security events by using graph-based retrieval to surface relevant context for each alert. In Fig.~\ref{fig:defengraph}, it presents the high-level system architecture.

When a SIEM alert is received, \textsc{DefenGraph} extracts relevant subgraphs from two complementary knowledge bases:

\begin{itemize}
    \item \textbf{Static Knowledge Graph (SKG)} encodes long-term infrastructure knowledge, typical attacker behaviors, associated SIEM alerts and defender response playbooks.

    \item \textbf{Dynamic Knowledge Graph (DKG)} captures time-evolving security events, attacker traces, and defender ticket actions recorded during cyber range exercises.
\end{itemize}

The retrieved relevant subgraph paths contextualize the alert in terms of:

\begin{itemize}
    \item Surrounding system structure (e.g., which users have access to this host?)
    \item Recent incidents (e.g., has this IP been flagged previously?)
    \item Prior defensive actions (e.g., what was done in similar cases?)
\end{itemize}

Because these subgraph paths are retrieved by a semantic-based method or neighbourhood connection (\S\ref{subsec:graph_retrieval}), they may contain noisy or overly broad information. To solve these problem, \textsc{DefenGraph} additionally introduces four key components:
\begin{itemize}
    \item \textbf{Graph-based retrieval process.} Utilises static and dynamic KGs to search for critical relational paths based on entity relationships.
    \item \textbf{LLM contextual filter.} Employs LLMs to evaluate and refine these relational paths based on contextually relevant SIEM alerts.
    \item \textbf{LLM reasoning re-ranker.} Based on the filtered contextual knowledge from the SKG, it employs the LLMs to score and reorder retrieved paths from DKG and ensure the most alert-relevant knowledge are highlighted.
    \item \textbf{Defense activity generation.} Translates those paths into actionable defense strategies, tailored to the alert and informed by both historical context and real-time insights.
\end{itemize}

\subsection{Knowledge Graph Construction}

In cybersecurity, KGs have emerged as a promising solution by integrating data from multiple domains with human-readable vocabularies. 

Transforming our ACDC datasets into a unified representation of entities and their relationships is challenging due to their diverse structure and context. Our dataset consists of:

\begin{itemize}
    \item \textbf{Unstructured:} Logs and chat messages requiring NLP for entity extraction.
    \item \textbf{Structured:} CSVs with system configurations and component details.
    \item \textbf{Programmatic:} Red Team scripts encoding attack actions and intent.
    \item \textbf{Semi-Structured:} JSON-formatted alerts and tickets mix structured fields with free text, complicating extraction and integration.
\end{itemize}

\subsubsection{Data Clean and Preprocessing}

The data cleaning process is essential for removing noise, reducing redundancy, and ensuring the consistency of logs, alerts, and other records within the ACDC dataset. The ACDC dataset contains noisy or irrelevant data, including system logs outside the events timeframes, and duplicate alerts. The following cleaning process ensures filtering noise and refining log data:

\begin{itemize}
    \item \textbf{Identify benign logs outside event scope.} 
    Logs falling outside the dates of ACDC events, such as benign activities unrelated to Red Team attacks, are identified and stored separately to avoid introducing noise.

    \item \textbf{Similarity-based filtering during event times.} During ACDC events, similarity filtering excludes logs matching known benign patterns, keeping only relevant event-related entries.
    
    \item \textbf{Resolution of duplicate Logs.} 
    Repeated logs, which commonly occur during brute-force or repetitive activities, are identified and resolved to maintain consistency and avoid redundancy.
\end{itemize}

\begin{figure*}
    \centering
    \includegraphics[width=0.9\linewidth]{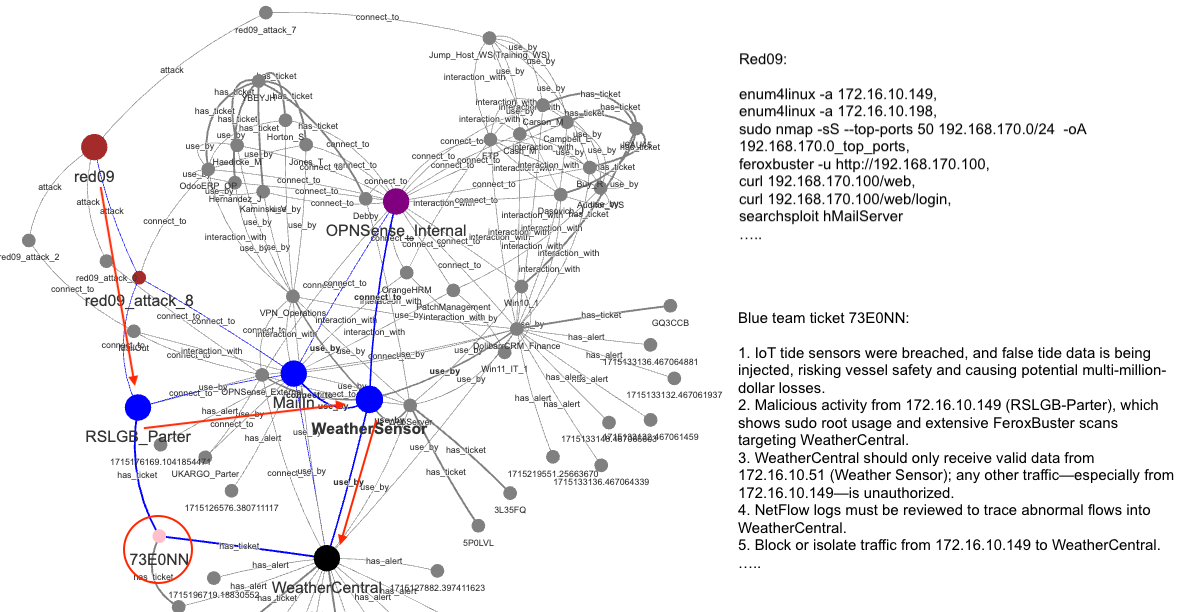}
    \caption{Red Team Attack Tracing. \textbf{Left} shows the attack trace from the red team. \textbf{Red Nodes} represent attacking actions performed by red team members. For example, the attack 8 (red09\_attack\_8) is carried out by the red team member 09. \textbf{Blue Nodes} and \textbf{Purple Nodes} indicate servers or firewalls. \textbf{Right} shows the attacking shell scripts used in red team attack 8, along with the corresponding defense ticket from the blue team. }
    \label{fig:attack_tracking}
\end{figure*}

\subsubsection{Named Entity Recognition (NER)}

Utilising spaCy~\cite{honnibal2020spacy}, identifies key entities such as server names, device identifiers, and attributes in free-text data. These entities form the nodes of the KG, with attributes like IP addresses, ports, and running services enhancing their specificity. 

\subsubsection{Red Team shell script grouping}

The Red Team shell scripts consist of timestamped commands. We use ChatGPT to analyze each command and explain its purpose and intent. We then group scripts by their objectives into logical clusters of related actions and link them to corresponding Wazuh alerts through similarity-based text searching. Finally, experts review the results.

\subsubsection{Relationship Extraction}

Relationships are extracted through a structured, multi-step process. Each stage helps identify important entities and the connections between them. This approach leads to a clear and well-structured knowledge graph that shows how the elements relate to one another:

\begin{itemize}
    \item \textbf{System architecture relationships.} Network and device relationships (e.g., firewalls, VLANs) are extracted from structured CSVs to define the system context. It helps discover the contextual insights and operational dynamics within the system. (see Supplementary V)

    \item \textbf{Entity-alert-logs mapping.} Entities detected by the NER process are identified in the Wazuh alerts. This step establishes impact relationships by connecting Wazuh alerts to the corresponding entities via those entity names. Similarly, each alert's timestamp allows it to be mapped to the relevant log. These matchings are further refined by using a text-based similarity method.

    \item \textbf{Alignment of Wazuh alerts and incident response tickets.} 
    Wazuh alerts and incident response tickets, containing the same entity names, will be applied to a similarity-based matching. A single incident response ticket may correspond to zero, one, or multiple Wazuh alerts. This is because one ticket might consolidate multiple related alerts if the Blue Team identifies them from the same issue. Alternatively, incident response tickets could be created proactively for precautionary measures, independent of any Wazuh alert. 

    \item \textbf{Linking red team scripts:} Grouped Red Team shell scripts are mapped to targeted entities or alerts. It reveals attack intent and scope.
\end{itemize}

\subsubsection{Static Knowledge Graph (SKG)} 
\begin{figure}
    \centering
    \includegraphics[width=0.95\linewidth]{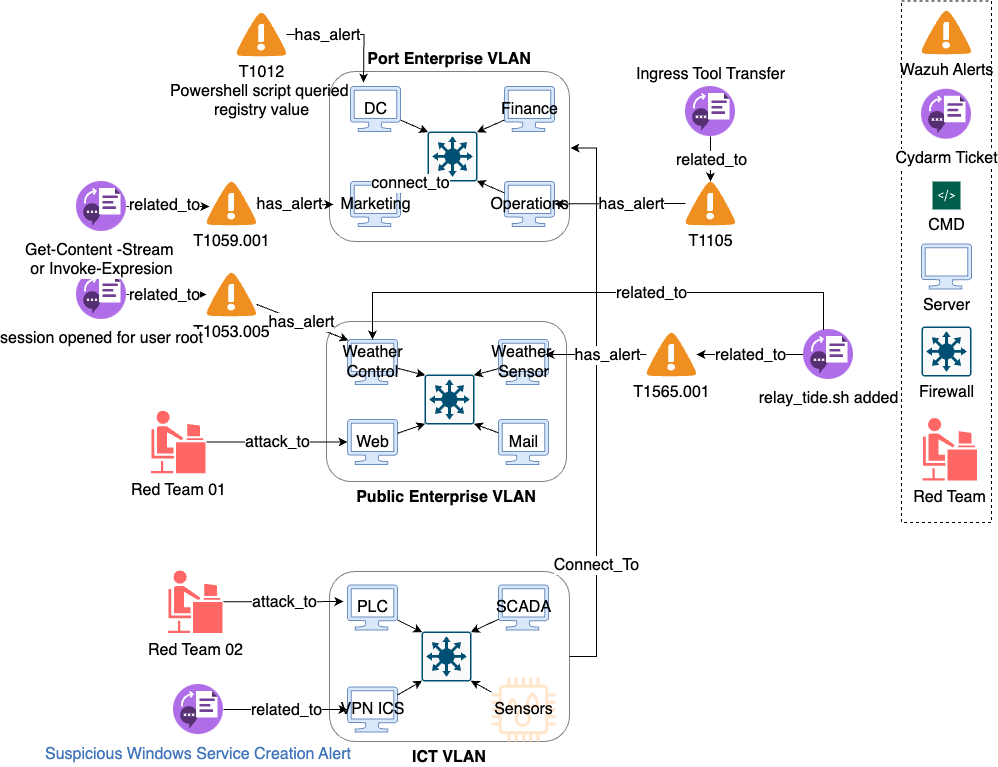}
    \caption{The ACDC Knowledge Graph example demonstrates the nodes and edges connection.}
    \label{fig:acdc_schema}
\end{figure}

The SKG encodes stable domain knowledge about system architecture, attacker behavior, and past defensive actions. It is curated from cyber range blueprints, attack emulation commands, attacker commands, and system documentation. Nodes represent entities such as hosts, users, tools, tactics, SIEM alerts, system logs, and incident response tickets. Edges capture relationships like “related\_to”, "has\_alert", “attack\_to”,  “responded\_with”, or "triggers", etc. An example of the SKG described above can be viewed in Fig.~\ref{fig:acdc_schema}.

This graph enables defenders and LLMs to trace prior attack-response chains across the cyber range timeline. For example, as shown in Fig.~\ref{fig:attack_tracking}, it links a successful compromise of the RSLGB\_Partner server by attacker red09 to a downstream breach of the tide sensor system, ultimately reflected in multiple correlated Blue Team tickets. By representing sequences of attacker actions and defender responses, the SKG supports reasoning over lateral movement, vulnerability exploitation, escalation, and recovery actions.

\subsubsection{Dynamic Knowledge Graph (DKG)} 

Unlike the SKG, which is derived from historical knowledge, the DKG always begins as a minimal system graph based on the system architecture shown in Supplementary V. The DKG grows continuously as new SIEM alerts, incident response tickets, logs and system events are observed. The same graph schema as the SKG with the addition of timestamps on every node. Each node represents a cyber artifact  (e.g., alert, ticket, IP, file), and edges encode temporal or semantic relationships (e.g., “related\_to”, "has\_alert", “attack\_to”,  “responded\_with”, or "triggers").

By capturing the order and co-occurrence of events, the DKG enables LLMs to reason about attack progression and recognize high-level intent. For instance, a sequence of alerts on the WeatherCentral server (starting with a web attack, followed by the addition of a suspicious script) can be linked to an incident ticket describing a tide sensor breach. This provides context for the LLM to identify the attacker’s goal: falsifying tide data to disrupt maritime navigation. While the full graph is retained during evaluation, practical deployment would require pruning to ensure tractability over time. These graphs provide a structured representation of defender’s operating environments and evolving attack behaviors.

\subsection{Graph-Based Retrieval Process}
\label{subsec:graph_retrieval}

The LLM generation is a sequence of tokens decoded step-by-step, which can accumulate errors and result in hallucinated reasoning paths and answers~\cite{nguyen-etal-2024-direct}. To address these challenges, we constrain the scope of the LLM’s context by anchoring it to knowledge derived from those retrieved graph paths. This restriction ensures that the LLM operates within a space of contextually relevant information, minimizing hallucinations~\cite{li2024subgraphrag}.  

\subsubsection{Input Preprocessing}
\label{sec_input_process}

\begin{figure}
    \centering
    \includegraphics[width=0.7\linewidth]{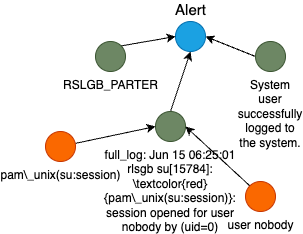}
    \caption{Wazuh alert graph query on the entity node RSLGB\_Parter. }
    \label{fig:rslgb_parter_alert_graph}
\end{figure}

The Graph-Based Retrieval Process (GRP) begins with input preprocessing. Each Wazuh alert $W_q$ is preprocessed to construct a query graph $G_q$. Entities in the graph represent the key components of the alert $W_q$, while relations between entities follow the schema shown in Fig.~\ref{fig:acdc_schema}. For example, for the Wazuh alert in Fig.~\ref{fig:rag_passage_example}, the root entity is ``RSLGB\_Parter'', and the node entity of its title is ``System user successfully logged into the system'', and the detail node entity is ``full\_log''. Additional attributes such as terms (``user nobody'', ``pam\_unix'') extracted from free text are linked to the detail node entity as the attribute entity, shown in Fig.~\ref{fig:rslgb_parter_alert_graph}.

We then apply the following retrieval strategies to obtain relevant subgraphs from the knowledge graph $G_B$ for the input Wazuh alert $W_q$ by using Semantic-based Triple Matching and Path Propagation Score:

\begin{itemize}
    \item \textbf{Scenario 1: Root node with similar alert. } 
    We initiate the graph search by matching the alert's agent name (entity ID) to the node ID in $G_B$. If the target entity in the Wazuh alert $W_q$ exists as a node in the above KGs base $G_B$, we then compare the properties of its associated alerts and tickets with those of the input alert $W_q$. When a similarity is found, this provides informative context and history for immediate decision-making. 

    \item \textbf{Scenario 2: Similar cases on different nodes.}  
    If $W_q$ represents a similar scenario to past cases, but on a different target entity, we traverse the properties of $G_q$ and search for similar properties within $G_B$. This reveals contextually or behaviorally similar incidents, even if the nodes differ (e.g., similar privilege escalations on different servers).

    \item \textbf{Scenario 3: No relevant match.}  
    If the retrieval process failed at the above two searches, the system defaults to extracting the subgraph of the target node from the system’s architectural graph to support informed defensive actions, such as assessing network configuration and overall architecture. While this does not directly resolve the alert, it provides valuable context, understanding the broader system landscape and devising appropriate defensive measures.
\end{itemize}

\subsubsection{Semantic-based Triple Matching (STM)} 

The semantic-based Triple matching is computed between Wazuh alert triples in $G_q$ and those in $G_B$. Leveraging pre-trained language models (PLMs), we utilise the text encoder to encode triples $T_q = \left\langle e_{q}, r_q, e^{’}_{q} \right\rangle$ from $G_q$ and $T_g = \left\langle e_{g}, r_g, e^{’}_{g} \right\rangle$ from $G_B$, where $e$ and $e^{’}$ are entities and $r$ is the relation. The PLM encodes these as:
\begin{equation}
h_q = \text{PLM}(T_q), \quad h_g = \text{PLM}(T_g).
\end{equation}

Inspired by the NSM model~\cite{he2021improving}, we apply projection layers to $h_q$ and $h_g$ to compute the semantic match vector $m^{t}_{\langle h_q, h_g \rangle}$ between a $G_q$ and a $G_B$ triple at step $t$:
\begin{equation}
m^{t}{\langle h_q, h^t_g \rangle} = \sigma \left( h_q W_q^{t} \cdot h^t_{g} W_g^{t} \right),
\label{eq:match_score}
\end{equation} 
where $W^{t}_q, W^{t}_g$ are parameters of the t-step projection layers. $\sigma$ is the sigmoid activation function.

\subsubsection{Path Propagation Score (PMC)} 
Based on the generated STM features, the PMC module computes path matching scores. For a given a $T_q$ and $G_B$, we initialize the STM scores as $M_t = \{m^{0}{\langle h_q, h^0_g \rangle}, m^{1}{(h_q, h^1_g)}, ..., m^{t}{(h_q, h^t_g)}\}$ in the \eqref{eq:match_score}. Due to computational constraints, only the top-$k$ elements of $M_t$ are retained for the subsequent propagation step. At step $t$, the propagation scores at the last step $S^{t-1}$ are used as weights to aggregate features from the next neighboring triples and update the match scores:
\begin{equation}
S^{k}_t(T_q, T^{t}_g) = S^{k}_{t-1}(T_q, T^{t-1}_g) \cdot m^{k}_t{\langle h_q, h^t_g \rangle}.
\end{equation}
After T-step iterations, we obtain the entire path semantic scores $S^{k}_t$, which indicate the likelihood of entities being answers to the alert $W_q$ and used in both retrieval and reasoning stages.

\subsubsection{SubGraphs Composition after PMC}

After the PMC search, we retrieve the top-$k$ matching paths linked to Wazuh alerts and then merge paths with the same root into subgraphs $G_{\text{sub}}^{L}$, where $L \leq k$ denotes the number of subgraphs with distinct roots.

To limit the final subgraphs, we sample 2-hop neighbor entities around root node in $G_{\text{sub}}^{L}$. This depth ensures the capture of surrounding contextual information of the root node and enhances the comprehensiveness and utility of the retrieved knowledge. The resulting subgraphs $G_{\text{sub}}^{L}$ are then used for downstream LLM filter and reasoning analysis. 

\subsection{LLM-based Contextual Filter}
\label{subsec:contextual_filter}

Due to the lack of global context in semantic retrieval, the resulting subgraphs often contain noisy and irrelevant information. Such noise can lead to incorrect or suboptimal choices, as highlighted by Zhang et al.~\cite{zhang-etal-2022-subgraph}. Inspired by LLM-CF~\cite{sun2024largelanguagemodelsenhanced}, LLM introduces global semantic awareness that traditional graph-based retrieval lacks. We apply LLM as a contextual filters to score key entities and critical paths, improving answer accuracy and coherence.

The contextual filter (CF) selects the top-$N$ paths $P$ according to the given Wazuh alert $W_q$, as given by \eqref{eq:p_topn}, that terminate with the tail entity within the alert $W_q$. The selected paths $P_{\text{top-N}}$ are then served for the LLM reasoning-based defense activity generation. 
\begin{equation}
P_{\text{top-N}} = \arg\max_{P_i \in \mathcal{P(G_{\text{sub}}^{L})}} S_{context}^{LLM}(P_i|W_q)
\label{eq:p_topn}
\end{equation}
where  $P_{\text{top-N}}$ represents the top-$N$ paths selected by the LLM,  $\mathcal{P}$ is the total paths in the subgraphs.

This filter approach not only enhances the accuracy by focusing on relevant subgraphs but also ensures that the process remains interpretable. By incorporating LLMs in the reasoning workflow, the proposed module effectively filters out noise and identifies the most meaningful connections and derives a refined subset from the previously retrieved subgraphs $G_{\text{sub}}^{L}$. 

\subsection{Temporal Weight Assignment} 

To retrieve time-sensitive knowledge from DKG, we propose a method that incorporates entity weighting based on timestamps, leveraging the structural and semantic strengths of both graph types. As described in the SKG section, SKG holds immutable foundational data, where temporal weighting is unnecessary but relational structure is preserved. In contrast, DKG captures time-sensitive information, with each entity node $e_i$ associated with a timestamp $t_i$. 

Each entity $e_i$ in the dynamic graph $G^{t}_{D}$ is assigned a temporal weight $w_t$ based on its timestamp $t_i$ and the query time $t_q$:
\begin{equation}
w_t = f(t_i, t_q) = exp(-\alpha|t_i - t_q|),
\end{equation}
where $\alpha$ is a decay factor that controls the importance of the recency. This ensures entities closer in time to the query are given higher importance.

\subsection{LLM Reasoning Rerank and Defense Activities Generation}
\label{subsec:reason_rerank}

The subgraphs composed of both static graph $G^S_{\text{sub}}$ and dynamic graph  $G^D_{\text{sub}}$, provide the LLM reasoning rerank module with foundational system knowledge and real-time contextual data. The module then examines $G^D_{\text{sub}}$ to identify and rerank reasoning paths $\mathcal{P}(G)$ based on their relevance to the Wazuh alert $W_q$ and experienced knowledge $G^S_{\text{sub}}$, using a prompt-based scoring mechanism. These paths, defined in \eqref{eq:p_topn2}, capture patterns and explanations informed by both historical and live system activity.
\begin{equation}
\mathcal{P}_{\text{rank-}N} = \arg\max_{P_i \in \mathcal{P}(G^D_{\text{sub}})}^{N} (S_{reason}^{LLM}(P_i \mid W_q, G^S_{\text{sub}}) \cdot W_t ).
\label{eq:p_topn2}
\end{equation}

Where $W_t$ is the temporal weight that penalises temporal distance from the query time. These high-scoring reasoning paths are subsequently synthesized into defense strategies via LLM-guided action generation. The resulting strategies are:

\begin{itemize}
    \item \textbf{Context-aware.} Informed by the entities and relationships directly relevant to the alert $W_q$ and its surrounding graph neighborhood in $G$.
    \item \textbf{Explainable.} Derived from interpretable reasoning chains $P_i$ that can be traced and audited by analysts.
    \item \textbf{Timely.} Weighted by temporal relevance from $G_d$, ensuring responsiveness to ongoing or emerging threats.
\end{itemize}

\section{Experiments}
\label{sec:exp}

We evalaute \textsc{DefenGraph} on the ACDC dataset in \S\ref{sec:acdc_dataset}. This complex environment enables realistic exercises where Blue Team defenders respond to Red Team attacks using Wazuh for threat detection and incident response systems for case management. 

To evaluate system performance, we use 32 of attacks–alerts-ticket triples to populate the SKG with historical attack–response context, while the remaining 24 attacks–alerts-ticket triples serve as the held-out evaluation set. Totally 56 attacks–alerts-ticket triples are used in the ACDC dataset. Importantly, the SKG is constructed using tickets from Red–Blue team exercises that are entirely disjoint from those used in the evaluation, ensuring disjoint training and test contexts. LLM prompts and details are provided in Supplementary IV.

We benchmark \textsc{DefenGraph} against four alternative approaches, including both non-graph-based and graph-based methods: LLM-Only, Hybrid RAG~\cite{yuan2024hybridragcomprehensiveenhancement}, GraphRAG~\cite{edge2024localglobalgraphrag},and \textsc{DefenGraph} (with SKG only and DKG+SKG). The evaluation is conducted across four LLMs (GPT-4o~\cite{DBLP:journals/corr/abs-2303-08774}, LLaMA 3.1~\cite{grattafiori2024llama3}, DeepSeek R1~\cite{deepseekai2025deepseekr1incentivizingreasoningcapability} and QWen3~\cite{qwen3}). We include LLM-Only and Hybrid RAG in our comparison even though they are not graph-based because these approaches represent the most common retrieval methods used in practice and demostrate graph-based design truly brings additional value.

Our experiments address the following research questions (RQs):  

\begin{itemize}
    \item \textbf{RQ1:} How effectively does \textsc{DefenGraph} generate accurate, context-specific defense actions?
    \item \textbf{RQ2:} How efficiently can the LLM identify relevant entities and reason over attack-response context?
    \item \textbf{RQ3:} How robust is \textsc{DefenGraph} in cold-start scenarios with no prior ticket history?
    \item \textbf{RQ4:} What is the real-time runtime performance of \textsc{DefenGraph} during live inference?
\end{itemize}

\subsection{Evaluation Metrics}
Traditional evaluation metrics such as HITs, precision, and recall are insufficient for assessing cyber defense recommendations, as they reduce assessment to binary relevance and overlook the depth and reasoning behind each decision. Blue team tickets are created manually, and they usually capture only the final investigation outcomes rather than the full investigative context. For example, as shown in the second part of Fig.~\ref{fig:rag_passage_example}, the intermediate investigation steps are absent from the ticket. In contrast, \textsc{DefenGraph} provides intermediate reasoning steps and explainable responses. Therefore, we aim to ensure that these reasoning steps align with the target entities referenced in the tickets and that the recommended defense actions correspond to the ticket outcomes.

To justify the richer intermediate reasoning steps and evaluate the precision and actionability of LLM's outputs, we introduce three custom metrics:

\begin{itemize}

\item \textbf{Reasoning analysis recall (RAR)} quantifies how well the LLM system identifies relevant attack targets within context. Higher RAR indicates more accurate alignment with the underlying threat logic.
\[
RAR = \frac{1}{\left\|P\right\|} \sum_{k=1}^{p} \frac{R_k}{T_k}
\]
where \( R_k \) is the number of correctly identified targets in the \( k \)-th ticket pair, and \( T_k \) is the total predicted targets.

\item \textbf{Ticket actions recall (TAR)} quantifies how many ground-truth defense actions from Blue Team tickets are recovered by the LLM. It quantifies the proportion of ground-truth defense actions fully addressed by the corresponding \textsc{DefenGraph}'s actions $S_k$.  

\[
TAR = \frac{1}{\left\| P \right\|}\sum_{k=1}^{p}\frac{S_k \cap G_k}{\left\| G_k \right\|}
\]

where $P$ is the pairs of the Wazuh alerts and the ground-truth tickets, $S_k$ represents the defense actions suggested in the $k$-th recommendation ticket, $G_k$ corresponds to the actions in the $k$-th ground-truth defense ticket, and $S_k \cap G_k$ indicates the overlap of actions between the recommended and ground-truth tickets.

\item \textbf{Ticket actions precision (TAP)} quantifies the proportion of suggested actions that are accurate relative to the total actions from \textsc{DefenGraph}. It provides a numerical measure of precision and shows how well the suggested actions match the correct defense actions. A high score means \textsc{DefenGraph} includes many of the actions found in the ground truth, while a low score means it suggests many actions that are not in the ground-truth records.
\[
TAP = \frac{1}{\left\|P\right\|} \sum_{k=1}^{p} \frac{|S_k \cap G_k|}{\left\|S_k\right\|}
\]
where \( S_k \) and \( G_k \) denote predicted and ground-truth action sets.

\item \textbf{Fault actions}: Actions that mistakenly impact benign services and are classified as  false positives.

\end{itemize}

\begin{table*}[t]
\scriptsize
\centering
\renewcommand{\arraystretch}{1.3}
\caption{
Performance comparison across LLM-only (No KB), Hybrid RAG, Static KG, and \textsc{DefenGraph} methods. \textbf{Identified Targets} are correctly inferred threat entities. \textbf{Total Actions} is the number of generated defense steps, and \textbf{Actions in BT Tickets} are those matching Blue Team responses. \textbf{Fault Actions} denote incorrect or harmful suggestions. ``\textbf{$\uparrow$}'' indicates higher is better; ``\textbf{$\downarrow$}'' lower is better.} 
\label{tab:metrics}
\begin{tabular}{|c|l|c|c|c|c|c|c|c|}
\hline
\multicolumn{1}{|c|}{\textbf{LLMs}}     & \multicolumn{1}{c|}{\textbf{Methods}} & \multicolumn{1}{c|}{\textbf{RAR}} & \multicolumn{1}{c|}{\textbf{TAR}} & \textbf{TAP}     & \multicolumn{1}{l|}{\textbf{\begin{tabular}[c]{@{}l@{}}Identified \\ Targets $\uparrow $\end{tabular}}} & \multicolumn{1}{l|}{\textbf{\begin{tabular}[c]{@{}l@{}}Total \\ Actions $\downarrow $\end{tabular}}} & \multicolumn{1}{l|}{\textbf{\begin{tabular}[c]{@{}l@{}}Actions in \\ BT Tickets $\uparrow $\end{tabular}}} & \multicolumn{1}{l|}{\textbf{\begin{tabular}[c]{@{}l@{}}Fault \\ Actions $\downarrow$\end{tabular}}} \\[1ex] 
\hline
\hline 
\multirow{5}{*}{\textbf{GPT-4o}}    & No KB (LLM Only) & 33.73\%   & 14.49\%   & 5.81\%    & 28    & 172   & 10  & \textbf{5}   \\
                                    & + Hybrid RAG & 33.65\%   & 20.29\%   & 8.19\%    & 27    & 171   & 14 & 6   \\
                                    & + GraphRAG (SKG) & 45.78\%   & 39.63\%   & 11.19\%    & 38    & \textbf{134}   & 15 & 8   \\
                                    \cline{2-9}
                                    & + DG (SKG + GRP + CF) & 61.45\%   & 52.17\%   & 24.49\%   & 51    & 147   & 36  & 8   \\
                                    & + DG (Full) & \textbf{73.49\%}  & \textbf{72.46\%}  & \textbf{29.24\%}   & \textbf{61}  &   171 & \textbf{50}  & 6  \\ \hline
\multirow{4}{*}{\textbf{LLaMA 3.1 253B}}    & + Hybrid RAG       & 12.05\%   & 8.71\%    & 4.76\%    & 10    & 126   & 6 & 6    \\ 
                                            & + GraphRAG (SKG) & 27.94\%   & 19.35\%   & 6.85\%    & 19    & 73   & 5 & 9   \\
                                            \cline{2-9}
                                            & + DG (SKG + GRP + CF)        & 46.99\%   & 42.03\%   & 29.01\%   & 39    & \textbf{100}  & 29 & 7   \\
                                            & + DG (Full)    & \textbf{61.45\%}   & \textbf{56.52\% }  & \textbf{36.11\%}   & \textbf{51}    & 108   & \textbf{39} & \textbf{5}   \\ \hline
\multirow{4}{*}{\textbf{DeepSeek R1 671B}}  & + Hybrid RAG   & 10.84\%   & 10.14\%   & 5.65\%    & 9     & 124   & 7 & \textbf{8} \\
                                            & + GraphRAG (SKG) & 38.23\%   & 25.81\%   & 7.36\%    & 26    & 95   & 7 & 14   \\
                                            \cline{2-9}
                                            & + DG (SKG + GRP + CF)    & 45.78\%   & 43.48\%   & 28.57\%   & 38    & \textbf{105}   & 30 & 13    \\
                                            & + DG (Full)        & \textbf{56.63\%}   & \textbf{59.42\%}   & \textbf{36.94\%}  & \textbf{47}    & 111   & \textbf{41} & 11    \\ \hline
\multirow{4}{*}{\textbf{QWen 3 235B}}    & + Hybrid RAG   & 24.1\%    & 11.59\%   & 6.72\%    & 20    & 119   & 8 & 6\\
                                        & + GraphRAG (SKG) & 35.29\%   & 29.03\%   & 7.26\%    & 24    & 124   & 9 & 8   \\
                                        \cline{2-9}
                                        & + DG (SKG + GRP + CF)    & 51.81\%   & 42.03\%   & 24.58\%   & 43    & \textbf{118}   & 29 & 5    \\
                                        & + DG (Full)        & \textbf{59.04\%}   & \textbf{46.38\%}   & \textbf{25.20\%}       & \textbf{49}    & 127 & \textbf{32} & \textbf{4} \\ 
                                        \hline
\end{tabular}
\end{table*}

\begin{figure}[t]
\scriptsize
\centering
\fbox{%
  \begin{minipage}{0.45\textwidth}
    \textbf{1. Over-reaction:}
    \begin{itemize}[leftmargin=*, itemsep=0pt]
        \item \textbf{Wazuh Alert:} Registry Key Integrity Checksum Changed
        \item \textbf{Suggested Action:} Isolate Jump\_Host\_WS from the network: \\
        \quad 1. \texttt{sudo iptables -I INPUT -s 192.168.30.74 -j DROP} \\
        \quad 2. \texttt{sudo iptables -I OUTPUT -d 192.168.30.74 -j DROP}
    \end{itemize}
    \rule{\linewidth}{0.2pt}
    \textbf{2. False positive:}
    \begin{itemize}[leftmargin=*, itemsep=0pt]
        \item \textbf{Wazuh Alert:} The following fatal alert was received: 70.
        \item \textbf{Suggested Action:} Temporarily block RDP to prevent possible exploitation: \\
        \quad \texttt{sudo firewall-cmd --add-rich-rule='rule family=ipv4 source address=192.168.174.115 service name=rdp reject'}
    \end{itemize}
  \end{minipage}
}
\caption{Examples of incorrect aggressive actions suggested by LLMs. Top: unnecessary isolation. Bottom: RDP blocked in response to benign SSL alert.}
\label{fig:false_actions_example}
\end{figure}

\subsection{Results}
We evaluate \textsc{DefenGraph} (DG) across four key questions, focusing on accuracy, reasoning efficiency, cold-start robustness, and real-time performance. As shown in Table~\ref{tab:metrics}, we compare five methods. \textbf{No KB} indicates that the LLM prompt includes no knowledge base and only the input alert. \textbf{Hybrid RAG} uses a text-only knowledge base, providing additional passages in the prompt. For \textbf{GraphRAG}, we use the same SKG as in \textsc{DefenGraph} as the knowledge base, but GraphRAG relies solely on semantic similarity–based retrieval method. \textbf{DG (SKG + GRP + CF)}, referred to as \textbf{DG (SKG)}, uses the SKG as the knowledge base along with our graph retrieval method (GRP) and LLM-based filter (CF). Finally, \textbf{DG (Full) }extends DG (SKG) by adding additional DKG data and a reasoning-based re-ranker.

\vspace{2mm}

\noindent \textbf{1. Defense Action Accuracy (RQ1).} 
Hybrid RAG yields the lowest TAR among knowledge base (KB) approaches, with its RAR slightly below even the LLM-only (No KB) baseline for GPT-4o (33.65\% vs. 33.73\%), likely due to noise introduced by similarity-based retrieval.

KG-based methods marginally outperformed  Hybrid RAG. GraphRAG achieves by nearly 25\% in TAR, benefiting from its structured representation of experiential attack–response knowledge. DG (SKG) even achieves 31\% more in TAR improvement than Hybrid RAG, reaching 42.03\% on Qwen3 and 52.17\% on GPT-4o. It also surpasses GraphRAG by 12\% because of our graph retrieval method and the LLM-based filter. \textsc{DefenGraph}-Full (DG-Full) boosts its performance further by integrating additional DKG with real-time Wazuh alerts, logs, and incident response tickets, and LLM-based reasoning re-rank. It achieves the highest performance across models, with TAR reaching 46.38\% on Qwen3 and 72.46\% on GPT-4o. On GPT-4o, this corresponds to a 20\% TAR gain and nearly 5\% TAP improvement over the DG (SKG).

While TAP lags slightly behind TAR, especially on GPT-4o and QWen3, this is expected: DG-Full generates additional, plausible investigative actions that extend beyond the concise entries recorded in tickets. You can clearly see the increased number of \textbf{Total Actions} generated by DG-Full is higher than the \textbf{Actions in BT Tickets}.
Therefore, DG-Full identifies more targets and better recovers ground-truth defense actions from incident response tickets than all baselines. Importantly, DG-Full’s fault action rate remains below 4\% of the total actions when evaluated using GPT-4o. \textbf{Fault Actions} fall into one of two categories: (1) over-reactions and (2) false positives. As illustrated in Fig.~\ref{fig:false_actions_example}, one example of over-reaction involves fully blocking a server, whereas a more appropriate response would have been rate-limiting or selectively filtering the offending IP. In a false positive case, the system misinterprets a benign TLS handshake failure ("The following fatal alert was received: 70") as an RDP-related threat, triggering an unnecessary block. These faults are easily caught by human defenders, reinforcing \textsc{DefenGraph}’s role as a support tool, which is not a standalone decision-maker.

\vspace{2mm}
\noindent \textbf{2. Reasoning Efficiency (RQ2).}
As shown in Table~\ref{tab:metrics}, Hybrid RAG retrieves the smallest portion of relevant knowledge, as evidenced by its low Reasoning Analysis Recall (RAR) signifying reduced alignment with ground-truth defense tickets. 
The GraphRAG and DG (SKG) perform better, leveraging structured context to improve reasoning coverage.

DG-Full consistently outperforms both baselines across all LLMs and metrics, achieving up to \textbf{40\%} higher RAR than RAG and \textbf{12\%} over Static KG on GPT-4o. The improvements are further reflected in the number of \textbf{Identified Targets} and \textbf{Actions in Blue Team Tickets}, demonstrating that dynamically enriching the KG with real-time context enables more accurate reasoning.

\begin{figure*}[t]
    \centering
    \subfigure[The LLM-based contextual filtering step removes a large amount of unnecessary information from the subgraph produced by the GRP method in the SKG. It retains only the pieces related to the task, reducing to a minimal subgraph for \textbf{DG (SKG)}.\label{fig:llm_filter_static}]{
        \includegraphics[width=0.9\linewidth]{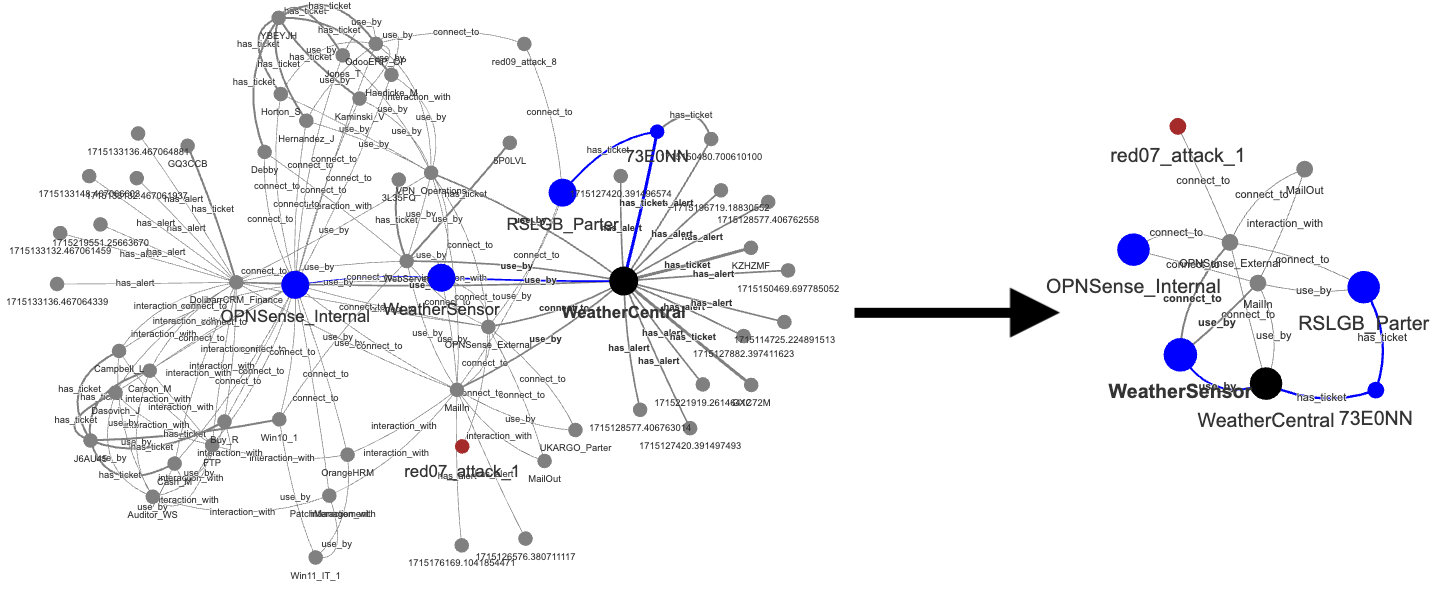}
    }

    \subfigure[The LLM-based contextual filtering step removes irrelevant information not only from the subgraph taken from the SKG but also from the subgraph extracted from the DKG. As a result, \textbf{DG-Full} (\textsc{DefenGraph} Full) can include the most up-to-date dynamic information, such as the node \textbf{RVNJL6}, and use it more effectively during analysis.\label{fig:llm_filter_dynamic}]{
        \includegraphics[width=0.9\linewidth]{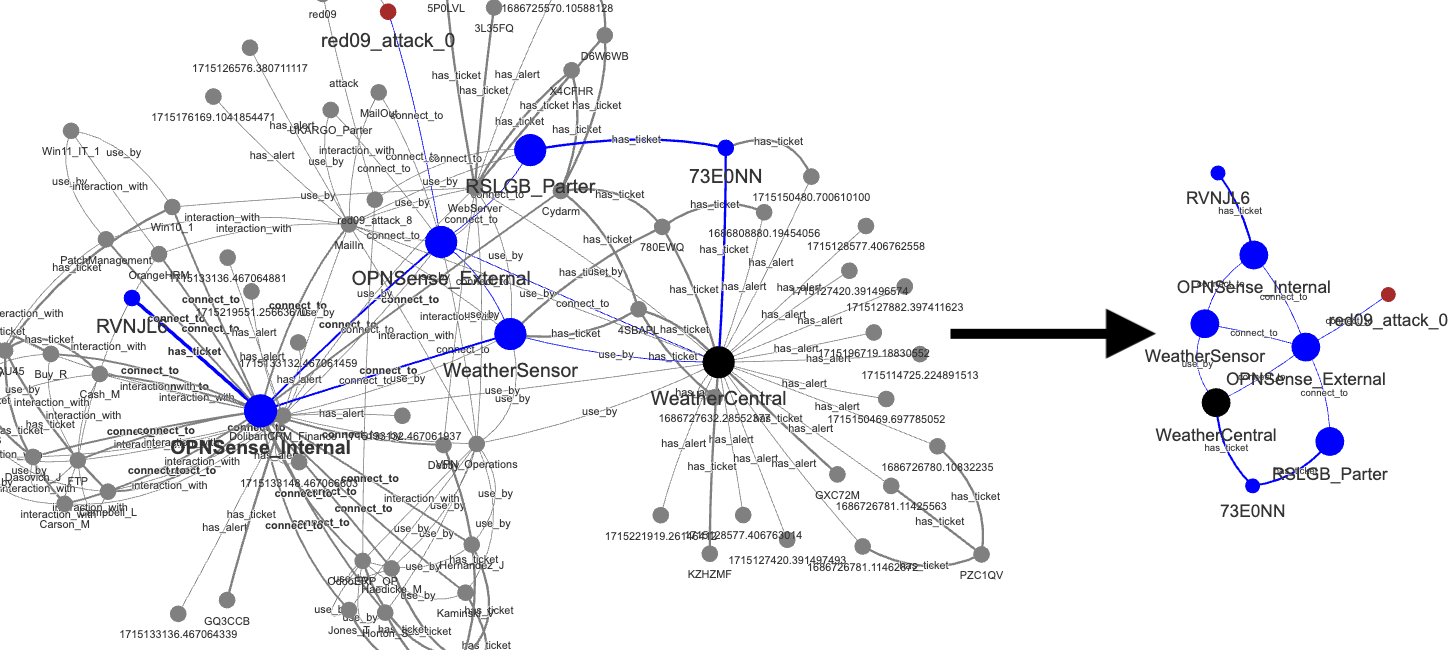}
    }

    \caption{Output comparison of LLM-based Contextual Filter between the DG (SKG) and the DG-Full. \textbf{Left}: the subgraph comes from the Graph-based Retrieval Process (GRP). \textbf{Right}: the smaller subgraph is the output produced by the LLM-based Contextual Filter. \textbf{Gray nodes} indicate information irrelevant to the input Wazuh alert. \textbf{Colored nodes} represent selected relevant information that also appears in the corresponding ground-truth ticket.}
    \label{fig:llm_filter_graph_demo}
\end{figure*}

\begin{figure}[t]
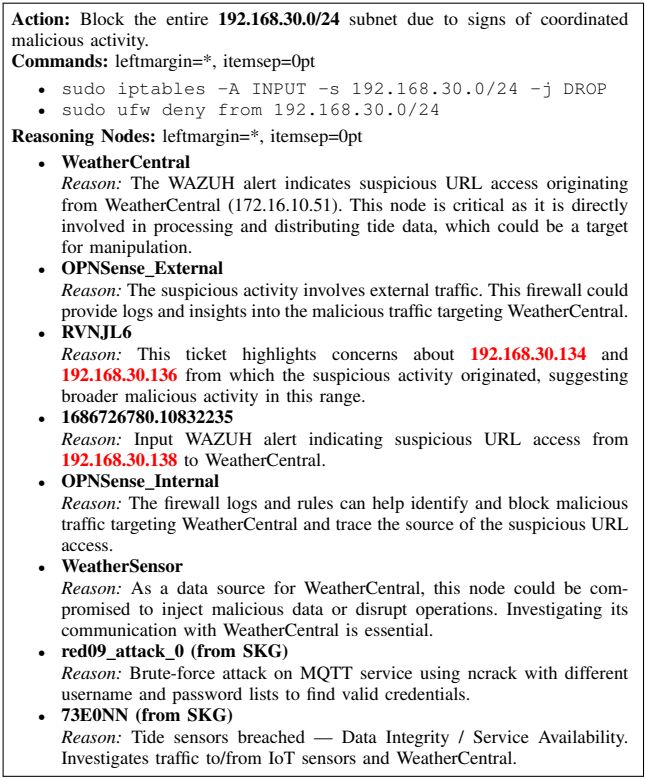

\scriptsize
\centering
\fbox{%
  \begin{minipage}{0.45\textwidth}
    \textbf{Action:} Block the entire \textbf{192.168.30.0/24} subnet due to signs of coordinated malicious activity. \\
    \textbf{Commands:}
    \begin{itemize}[leftmargin=*, itemsep=0pt]
        \item \texttt{sudo iptables -A INPUT -s 192.168.30.0/24 -j DROP}
        \item \texttt{sudo ufw deny from 192.168.30.0/24}
    \end{itemize}

    \textbf{Reasoning Nodes:}
    \begin{itemize}[leftmargin=*, itemsep=0pt]
    \item \textbf{WeatherCentral} \\
    \textit{Reason:} The WAZUH alert indicates suspicious URL access originating from WeatherCentral (172.16.10.51). This node is critical as it is directly involved in processing and distributing tide data, which could be a target for manipulation.
    
    \item \textbf{OPNSense\_External} \\
    \textit{Reason:} The suspicious activity involves external traffic. This firewall could provide logs and insights into the malicious traffic targeting WeatherCentral.
    
    \item \textbf{RVNJL6} \\
    \textit{Reason:} This ticket highlights concerns about \textcolor{red}{\textbf{192.168.30.134}} and \textcolor{red}{\textbf{192.168.30.136}} from which the suspicious activity originated, suggesting broader malicious activity in this range.
    
    \item \textbf{1686726780.10832235} \\
    \textit{Reason:} Input WAZUH alert indicating suspicious URL access from \textcolor{red}{\textbf{192.168.30.138}} to WeatherCentral. 
    
    \item \textbf{OPNSense\_Internal} \\
    \textit{Reason:} The firewall logs and rules can help identify and block malicious traffic targeting WeatherCentral and trace the source of the suspicious URL access.
    
    \item \textbf{WeatherSensor} \\
    \textit{Reason:} As a data source for WeatherCentral, this node could be compromised to inject malicious data or disrupt operations. Investigating its communication with WeatherCentral is essential.
    
    \item \textbf{red09\_attack\_0 (from SKG)} \\
    \textit{Reason:} Brute-force attack on MQTT service using ncrack with different username and password lists to find valid credentials. 
    
    \item \textbf{73E0NN (from SKG)} \\
    \textit{Reason:} Tide sensors breached — Data Integrity / Service Availability. Investigates traffic to/from IoT sensors and WeatherCentral.     
    \end{itemize}

  \end{minipage}
}
\caption{Example generated action generated by Full \textsc{DefenGraph} derived from the static-dynamic KG with LLM-based reasoning context in Fig.~\ref{fig:llm_filter_graph_demo}(b).}
\label{fig:llm_action_dynamic}
\end{figure}

\begin{figure}[t]
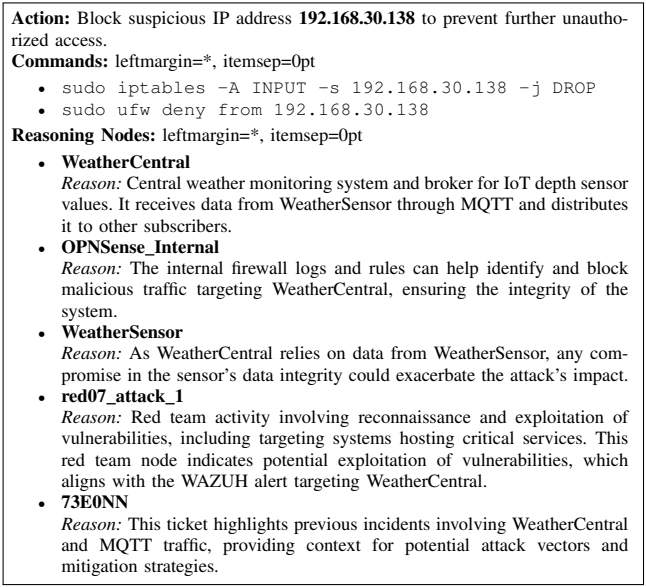

\scriptsize
\centering
\fbox{%
  \begin{minipage}{0.45\textwidth}
    \textbf{Action:} Block suspicious IP address \textbf{192.168.30.138} to prevent further unauthorized access. \\
    \textbf{Commands:}
    \begin{itemize}[leftmargin=*, itemsep=0pt]
        \item \texttt{sudo iptables -A INPUT -s 192.168.30.138 -j DROP}
        \item \texttt{sudo ufw deny from 192.168.30.138}
    \end{itemize}

    \textbf{Reasoning Nodes:}
    \begin{itemize}[leftmargin=*, itemsep=0pt]
    \item \textbf{WeatherCentral} \\
    \textit{Reason:} Central weather monitoring system and broker for IoT depth sensor values. It receives data from WeatherSensor through MQTT and distributes it to other subscribers. 
    
    \item \textbf{OPNSense\_Internal} \\
    \textit{Reason:} The internal firewall logs and rules can help identify and block malicious traffic targeting WeatherCentral, ensuring the integrity of the system.
    
    \item \textbf{WeatherSensor} \\
    \textit{Reason:} As WeatherCentral relies on data from WeatherSensor, any compromise in the sensor's data integrity could exacerbate the attack's impact.
    
    \item \textbf{red07\_attack\_1} \\
    \textit{Reason:} Red team activity involving reconnaissance and exploitation of vulnerabilities, including targeting systems hosting critical services. This red team node indicates potential exploitation of vulnerabilities, which aligns with the WAZUH alert targeting WeatherCentral.
    
    \item \textbf{73E0NN} \\
    \textit{Reason:} This ticket highlights previous incidents involving WeatherCentral and MQTT traffic, providing context for potential attack vectors and mitigation strategies.
    
    \end{itemize}

  \end{minipage}
}
\caption{Example generated response by DG (SKG) based on static KG with LLM-based reasoning context in Fig.~\ref{fig:llm_filter_graph_demo}(a).}
\label{fig:llm_action_static}
\end{figure}

Fig.~\ref{fig:llm_filter_graph_demo} illustrates a case study comparing subgraphs retrieved by the SKG-based \textsc{DefenGraph} (a) and full \textsc{DefenGraph} (b), based on the same Wazuh alert (see Supplementary III.A). The corresponding ground-truth ticket is shown in Supplementary III.B.2, and \textsc{DefenGraph}’s output is in Supplementary III.B.3. The subgraph includes additional nodes such as \textbf{RVNJL6} and \textbf{OPNSense\_External}, selected by the LLM due to a real-time alert and ticket context. For example, \textsc{DefenGraph} say of ticket \textbf{RVNJL6}: \textit{This incident response ticket raises concerns about the subnet (192.168.30.0/24) from which suspicious activity originated, indicating potential broader malicious behavior within that range.} This reasoning leads to a subnet-wide block recommendation (Fig.~\ref{fig:llm_action_dynamic}), aligned with the ground-truth response. In contrast, the SKG-based \textsc{DefenGraph}, lacking real-time incident data, retrieves a narrower subgraph and recommends blocking only a single IP \textbf{192.168.30.138}, as shown in Figs.~\ref{fig:llm_action_dynamic} and~\ref{fig:llm_action_static}. This highlights \textsc{DefenGraph}’s improved situational awareness and alignment with human-authored defensive actions.


\vspace{2mm}
\noindent \textbf{3. Cold Start Generalization (RQ3).}
In our main evaluation (Table~\ref{tab:metrics}), the dynamic KG accumulates knowledge across three Red–Blue team events, reflecting a realistic deployment where historical context builds over time.

To simulate cold-start conditions, we reset the dynamic KG before each event (Table~\ref{tab_cold_start}), removing prior ticket knowledge. In this setting, \textsc{DefenGraph} relies solely on the static KG and current alerts.

While RAR remains relatively strong (e.g., 58.26\% for GPT-4o), TAR and TAP drop noticeably, showing the system struggles to recover precise, ticket-aligned actions without dynamic context. These results suggest that overly short dynamic memory hinders performance. Retaining a richer dynamic KG across events or further into the past enables more accurate and complete defense recommendations.

\begin{table}[t]
\caption{Cold Start Performance of \textsc{DefenGraph} (DG) across different LLMs.}
\label{tab_cold_start}
\scriptsize
\centering
\renewcommand{\arraystretch}{1.2}
\begin{tabular}{|c|c|c|c|}
\hline
\textbf{Method} & \textbf{RAR} & \textbf{TAR} & \textbf{TAP} \\ 
\hline
\hline
\textbf{GPT-4o} + DG & 58.26\% & 37.43\% & 15.25\% \\ 
\hline
\textbf{LLaMA 3.1 253B} + DG & 42.21\% & 32.22\% & 17.37\%  \\ 
\hline
\textbf{DeepSeek R1 671B} + DG & 41.23\% & 33.11\% & 18.94\%  \\ 
\hline
\textbf{QWen 3 235B} + DG & 46.01\% & 32.34\% & 12.60\% \\ 
\hline
\end{tabular}
\end{table}


\vspace{2mm}
\noindent \textbf{4. Execution Time and Real-Time Performance (RQ4).} 
\textsc{DefenGraph} involves two types of computation: local processing and remote LLM calls. The local part includes STM, PMC, and subgraph composition in \S\ref{subsec:graph_retrieval}. For a query graph $G_q$ with $|q|$ edges and a knowledge-base graph $G_b$ with $|T|$ edges:

\begin{itemize}
\item The STM and PMC path-scoring steps require $O(N_q + N_g + kT)$ where $N_q$ and $N_g$ are the numbers of nodes in the query and background graphs, and $kT$ is the cost of keeping the Top-K paths over T propagation steps.
\item Subgraph composition takes $O(L^2)$, where $L \leq KL$ is the number of Top-K paths kept.
\end{itemize}

Thus, the total local computation complexity is $O(N + kT + L^2)$ which shows that the local method scales well in practice.

The remote LLM calls include LLM-based contextual filter, LLM reasoning rerank and defense activities generation. With GPT-4o, the total LLM API calls run ranging from 26.18 to 44.41 seconds. The LLM reasoning rerank and the defense activities generation costs around 21.16 seconds. We report them together because they run consecutively as one process. 
Therefore, \textsc{DefenGraph} invokes the LLM only three times per run, minimizing latency caused by LLM computation and communication. This design follows a similar two-stage approach as Think-Then-React~\cite{tan2025thinkthenreactunconstrainedhumanactiontoreaction}, which separates reasoning and generation to balance performance with efficiency. 
To further reduce the LLM API latency, local deployment or parallelization (multiple threads/processes) can support large-scale alert handling.

\subsection{Ablation Study}

Our approach includes several key components:
\begin{itemize}
    \item An experience-based knowledge graph (SKG) and a temporal knowledge graph (DKG) for capturing both long-term knowledge and real-time events;
    \item A graph-based retrieval process with contextual filtering to extract accurate, high-quality information, especially from the experience knowledge;
    \item A reasoning-based re-ranking step for generating the final defense actions from the processed information.
\end{itemize}
Based on these components, we conducted an ablation study to evaluate the contribution of each part.

\subsubsection{Static KG Only Without GRP or LLM-CF}
To understand how well our knowledge graph-based methods work in cyber-defense, Table \ref{tab:metrics} shows a clear pattern: no matter whether we use GraphRAG or DGs, all graph-based approaches perform noticeably better than traditional RAG across every model. For example, GraphRAG without our additional refinement methods scores much higher than HyberRAG. 

This advantage holds even in extreme cold-start situations where very little or no prior knowledge is available. Table \ref{tab_cold_start} confirms that graph-based methods still outperform traditional RAG in these challenging scenarios. In other words, graphs not only boost performance in normal settings but also remain more reliable when the system has almost no initial information to work with.

\begin{figure}
    \centering
    \includegraphics[width=0.95\linewidth]{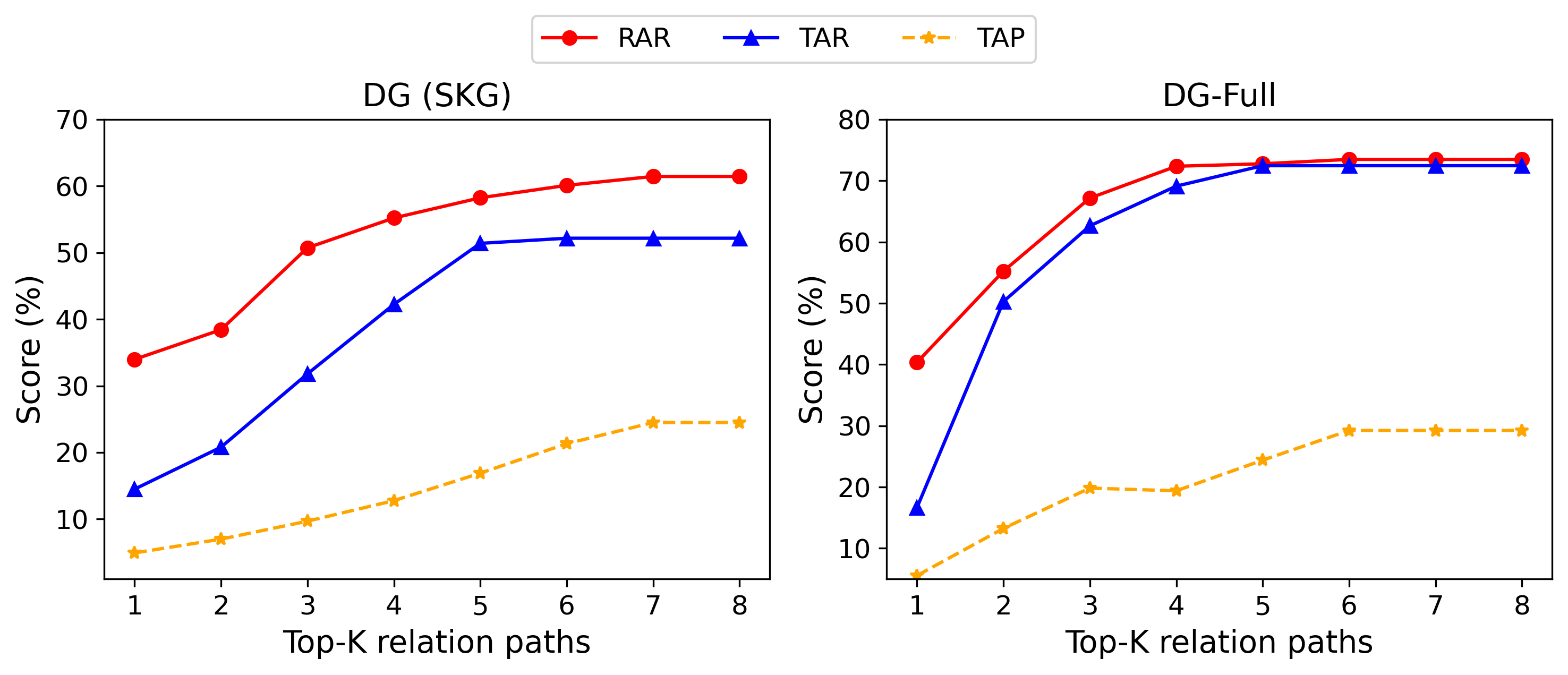}
    \caption{Demonstration of Top-K generated relation paths (\S\ref{subsec:contextual_filter}) and relevant scores: RAR, TAR and TAP. Red line denote the defence targets coverage of the retrieved paths. The other two lines denote the defence activities coverage of the retrieved paths.}
    \label{fig:topk_results}
\end{figure}

\begin{table}[t]
\centering
\renewcommand{\arraystretch}{1.3}
\caption{Performance metrics of GPT‑4o \textsc{DefenGraph} using Static KG (SKG), Dynamic KG (DKG), and Static-Dynamic KGs (DG-Full).}
\label{tab:metrics_ablation}
\begin{tabular}{|c|c|c|c|}
\hline
\textbf{Method} & \textbf{RAR} & \textbf{TAR} & \textbf{TAP} \\ 
\hline
\hline 
\textbf{DG (SKG Only)} & 61.45\% & 52.17\% & 24.49\% \\ 
\hline
\textbf{DG (DKG Only)} & 59.72\% & 47.28\% & 23.57\% \\ 
\hline
\textbf{DG-Full} & 73.49\% & 72.46\% & 29.24\% \\ 
\hline
\end{tabular}
\end{table}

\begin{table*}[t]
\centering
\caption{Comparison of KG-based cybersecurity frameworks with emphasis on actionable cyber defense support.}
\label{tab:kg_comparison}
\resizebox{\linewidth}{!}{
\begin{tabular}{l l l l c c}
\toprule
\textbf{Work} &
\textbf{Primary goal} &
\textbf{Primary output} &
\textbf{Main evidence source} &
\textbf{Temporal events} &
\textbf{Generates defense actions} \\
\midrule
Agrawal et al.~\cite{agrawal2024cyberq}
& Cyber reasoning
& Reasoning/analytics results
& Heterogeneous cyber data
& -- & --  \\

Qiao et al.~\cite{garrido2021machine}
& Intrusion detection
& Predicted links / alerts
& Network and security logs
& -- & --  \\

Zhu et al.~\cite{8600108}
& Attack source tracing
& Attack paths / correlations
& Space--ground network data
& -- & --  \\

CSKG4APT~\cite{9834133}
& APT analysis
& Queryable APT KG
& APT intelligence
& -- & -- \\

CTINexus~\cite{cheng2025ctinexusautomaticcyberthreat}
& CTI automation
& Extracted entities/relations
& CTI reports
& -- & --  \\

AttacKG+~\cite{ZHANG2025104220}
& Threat actor modeling
& Structured attacker behaviors
& CTI text reports
& -- & --  \\

Fieblinger et al.~\cite{fieblinger2024actionablecyberthreatintelligence}
& Actionable CTI
& Linked CTI knowledge for analysts
& Unstructured CTI
& -- & --  \\

KG-IBL~\cite{10720079}
& Few-shot learning
& Incremental model updates
& IIoT signal data
& -- & --  \\

Wrongdoing Monitor~\cite{9830760}
& Behavioral anomaly detection
& Behavioral anomaly indicators
& Behavioral event data
& -- & --  \\

K-GetNID~\cite{10.1109/TIFS.2024.3431932}
& Network intrusion detection
& Early/transferable detection
& Network flow data
& \cmark & -- \\

\midrule
\textsc{DefenGraph} \textbf{(Ours)}
& \textbf{Cyber defense analysis}
& \textbf{Ticket-aligned mitigation steps + reasoning}
& \textbf{Real-world IT/OT defense data}
& \cmark & \cmark  \\
\bottomrule
\end{tabular}
}
\end{table*}

\subsubsection{Static KG With GRP And LLM-based CF Without Dynamic KG}

To understand how the extra refinement methods (GRP and the LLM-based CF) work when using only the static knowledge graph (SKG), we removed both the dynamic knowledge graph and the LLM-based reasoning reranker described in \S\ref{subsec:reason_rerank}. In this setting, \textsc{DefenGraph} (SKG) relies only on SKG together with GRP and the LLM-based CF. Because it no longer receives real-time information, the defense activities recommendations leaned toward experienced or outdated blue team tickets and reduced the relevance of its responses to ongoing threats.

Table~\ref{tab:metrics} shows that performance improves when GRP and CF are added, as seen by comparing GraphRAG (SKG) with DG (SKG + GRP + CF). For GPT-4o, TAR increases by 12.54\%, TAP increases by 13.3\%, and RAR increases by 15.67\%. Across all LLMs, the smallest gain in RAR is 7.55\% on DeepSeek, the smallest gain in TAR is 13\% on QWen, and and the smallest gain in TAP is 17.32\%, also on QWen. 

Moreover, removing DKG shows a clear decrease in performance. For GPT-4o, TAR drops by 20.29\%, TAP drops by 4.75\%, and RAR decreases by 12.04\%. These results show that the DKG not only improves the model’s ability to adapt to new threats but also strengthens its overall reasoning quality.

\subsubsection{Dynamic KG With LLM-based Reasoning Rerank Without Static KG}

The static KG serves as the backbone of the reasoning process by providing well-structured, domain-specific information that enhances the LLM’s ability to comprehend and address cybersecurity threats. The removal of the static KG resulted in a significant performance decline across all evaluation metrics, most notably a 25\% drop in TAR as shown in Table~\ref{tab:metrics_ablation}. This highlights the essential role of structured historical knowledge in generating effective and context-aware defense actions.

\subsubsection{Impact of LLM-based Contextual Filter on Defence Activity Coverage}

To evaluate the faithfulness of the Top-K relation paths, we illustrate the quantitative results in Fig~\ref{fig:topk_results}. In experiments, we sweep the number of top-K relation paths generated by LLM-based Contextual Filter. From results, we can see that the number of retrieved relation paths (Top-K) influences three coverage metrics: RAR, TAR and TAP across both DG (SKG) and DG-Full settings. As K increases, all three metrics consistently improve, which demonstrates that retrieving more relation paths enriches the contextual evidence used by the model. RAR increases quickly and reaches higher values as more paths are added. TAR also becomes higher as K grows, especially in the DG-Full setting. TAP increases more slowly due to adding more relation paths also introduces more actions. Some of paths may be noise and reduce the contribution to precision, but the score still shows an overall upward trend. Overall, the results show that using more relation paths helps the model cover more defence information, and the improvements become smaller once K is around six to eight.

\section{Related Work}
\label{sec:rw}

Table~\ref{tab:kg_comparison} summarizes a qualitative comparison between \textsc{DefenGraph} and representative KG-based cybersecurity frameworks, with an emphasis on actionable defender support. Prior work has established that knowledge graphs provide structured representations of cyber entities and relations that improve reasoning, detection, and intelligence fusion across heterogeneous security artifacts~\cite{agrawal2024cyberq, 10.1049/cmu2.12736}. However, as reflected in Table~\ref{tab:kg_comparison}, most existing systems primarily target upstream tasks---such as cyber reasoning outputs~\cite{agrawal2024cyberq}, intrusion alerts via link prediction~\cite{garrido2021machine}, or attack-source tracing through inferred paths and correlations~\cite{8600108}, rather than producing defender-ready mitigation actions.

A second line of work focuses on structuring threat intelligence into queryable KGs. CSKG4APT~\cite{9834133} consolidates heterogeneous APT information into a unified, queryable knowledge graph to support analyst investigation and planning. CTINexus~\cite{cheng2025ctinexusautomaticcyberthreat} and AttacKG+~\cite{ZHANG2025104220} use LLM-assisted extraction pipelines to identify entities and relations from CTI reports and construct threat intelligence KGs with reduced annotation burden, while Fieblinger et al.~\cite{fieblinger2024actionablecyberthreatintelligence} connects unstructured CTI to linked knowledge to aid analyst workflows. These approaches largely operate over CTI-centric evidence and typically output structured knowledge (entities, relations, behaviors, or linked intelligence) rather than incident-specific defense actions.

Beyond CTI, KGs have been used for learning and detection in specialized security domains. KG-IBL~\cite{10720079} leverages a KG to store historical relationships and support incremental updates for few-shot emitter identification in IIoT settings, improving sample efficiency without full retraining. Wrongdoing Monitor~\cite{9830760} models event- and property-level behavioral associations to detect anomalies across scenarios, outputting behavioral anomaly indicators rather than mitigation steps. K-GetNID~\cite{10.1109/TIFS.2024.3431932} is the closest in spirit to temporal modeling, constructing a heterogeneous temporal graph for early and transferable network intrusion detection; nevertheless, its primary output remains detection rather than prescriptive defensive actions.

\smallskip
\noindent\textbf{Ours.}
In contrast to prior KG-based cybersecurity frameworks that predominantly produce reasoning results, alerts, attack paths, or structured threat intelligence, \textsc{DefenGraph} targets \textit{cyber defense analysis} and directly outputs \textit{ticket-aligned mitigation steps with supporting reasoning} grounded in real-world IT/OT operational artifacts (alerts paired with defender tickets). As highlighted in Table~\ref{tab:kg_comparison}, \textsc{DefenGraph} explicitly incorporates temporal incident evidence and is designed to generate actionable defense recommendations, bridging the gap between KG-enabled understanding and operationally useful response guidance.

\section{Conclusion}
\label{sec:conclusion}

This paper introduced \textsc{DefenGraph}, an LLM-driven cyber incident assistant that grounds recommendations in a dual-layer Static–Dynamic Knowledge Graph, combining long-term domain knowledge with evolving incident evidence. Given an incident query, \textsc{DefenGraph} retrieves multi-hop graph paths, filters irrelevant context, and re-ranks candidates with reasoning signals to surface the most defensible actions with stronger contextual and temporal fidelity. Evaluated on heterogeneous security artifacts collected from live Red vs.\ Blue cyber range exercises on critical-infrastructure scenarios, \textsc{DefenGraph} consistently outperformed strong baselines across four LLMs: on GPT-4o it raised reasoning recall from 61.45\% to 73.49\% and ticket-action recall from 52.17\% to 72.46\% (precision 24.49\%$\rightarrow$29.24\%), with similar gains on LLaMA-3, DeepSeek-R1, and QWen-3, and it surfaced up to 50 correct defense actions versus 36 for the next-best method, setting a new state-of-the-art for grounded, temporally aware cybersecurity decision support. 

\bibliographystyle{IEEEtran}
\bibliography{base_new}

\clearpage

\clearpage

\end{document}